\documentclass[conference]{IEEEtran}
\IEEEoverridecommandlockouts
\usepackage{cite}
\usepackage{amsmath,amssymb,amsfonts}
\usepackage{algorithmic}
\usepackage{graphicx}
\usepackage{textcomp}
\usepackage{xcolor}
\usepackage{rotating}
\usepackage{multirow}
\usepackage{color,soul}
\usepackage[inline]{enumitem}
\usepackage{glossaries}
\usepackage{balance}
\usepackage{hyperref}
\usepackage[inline]{enumitem}
\def\BibTeX{{\rm B\kern-.05em{\sc i\kern-.025em b}\kern-.08em
    T\kern-.1667em\lower.7ex\hbox{E}\kern-.125emX}}
\usepackage{todonotes}

\makeatletter
\newcommand{\linebreakand}{%
  \end{@IEEEauthorhalign}
  \hfill\mbox{}\par
  \mbox{}\hfill\begin{@IEEEauthorhalign}
}
\makeatother

\begin{document}

\title{Towards Developing Resilient and Service-oriented Mission-critical Systems}

\author{\IEEEauthorblockN{Do\u{g}analp Ergen\c{c}, Cornelia Br{\"u}lhart, Mathias Fischer}
\IEEEauthorblockA{\textit{University of Hamburg}, Germany \\
name.surname@uni-hamburg.de}
}

\maketitle
\begin{abstract}
Mission-critical systems~(MCSs) have embraced new design paradigms such as service-oriented architecture~(SOA) and IEEE 802.1 Time-sensitive Networking~(TSN). These approaches tackle the static and closed-loop design and configuration of MCSs to address their strict performance and resilience requirements. While SOA enables the dynamic placement of critical services over virtualized hardware, TSN provides several protocols to establish deterministic communication over standard Ethernet equipment.  
This paper presents a prototype combining SOA and TSN to design flexible and fault-tolerant MCSs. It demonstrates the benefits of dynamic service migration and time-sensitive redundancy protocols to increase the resilience of MCSs against node and link failures, respectively. Moreover, it presents additional advanced functionalities like optimal service distribution and security monitoring for new TSN protocols.
\end{abstract}


\section{Introduction}

Mission-critical systems~(MCSs) such as automobiles and avionics should be resilient against faults and attacks while also fulfilling strict quality of service~(QoS) requirements. They feature tightly coupled hardware and software modules and a closed-loop architecture, which strictly limits modifications and updates in their design and (re-)configuration. As a result, these systems are usually static and do not support dynamic adaptation, which can help to recover from faults and attacks. Therefore, MCSs have adopted new paradigms such as service-oriented architecture~(SOA) and IEEE 802.1 Time-sensitive Networking~(TSN), enabling flexible configuration and advanced resilience countermeasures.

SOA is a network softwarization approach that enables developing modular services, which can be dynamically distributed over virtualized systems~\cite{Heiser2011, Sim2018, Villaneueva2021}.
Another emerging technology, IEEE 802.1 TSN, provides networking protocols for time-sensitive communication over standard Ethernet equipment~\cite{ieee-tsn}. It also offers policing, filtering, and redundancy mechanisms for the safety and security of MCSs, as well as reduces their deployment and maintenance costs.
Nevertheless, implementing novel technologies poses additional challenges, as they introduce additional complexity and potentially new safety and security threats. 

In this paper, we develop a prototype combining SOA and IEEE 802.1 TSN protocols to investigate the novel design artifacts of next-generation MCSs. To the best of our knowledge, this is the first prototype combining these technologies and primarily focusing on their benefits for design flexibility and fault tolerance. Then, we demonstrate several scenarios regarding the optimal configuration, and the secure and fault-tolerant orchestration of this prototype. Our contributions are listed as follows. 
\begin{itemize}[leftmargin=*]
\item We have implemented virtualized system nodes and a supervisor capable of service orchestration and failover mechanisms to design a fault-tolerant SOA for MCSs.
\item We demonstrate the benefits of SOA and TSN against node and link failures.
\item We also introduce advanced features such as resource-optimal service distribution, status monitoring of services, and security monitoring for TSN traffic. 
\end{itemize}
In the following, Section~\ref{sec:architecture} presents an overview of our system architecture enabling SOA. Section~\ref{sec:prototype} presents our prototype and demonstration scenarios. Section~\ref{sec:conclusion} concludes the paper.

\section{System Architecture} \label{sec:architecture}
This section gives an overview of our system architecture, which constitutes a service-oriented and time-sensitive network. It also presents the primary components we implemented and describes their interconnectivity.

\subsection{System Overview}

\begin{figure}[ht!]
\centering
\includegraphics[width=\linewidth]{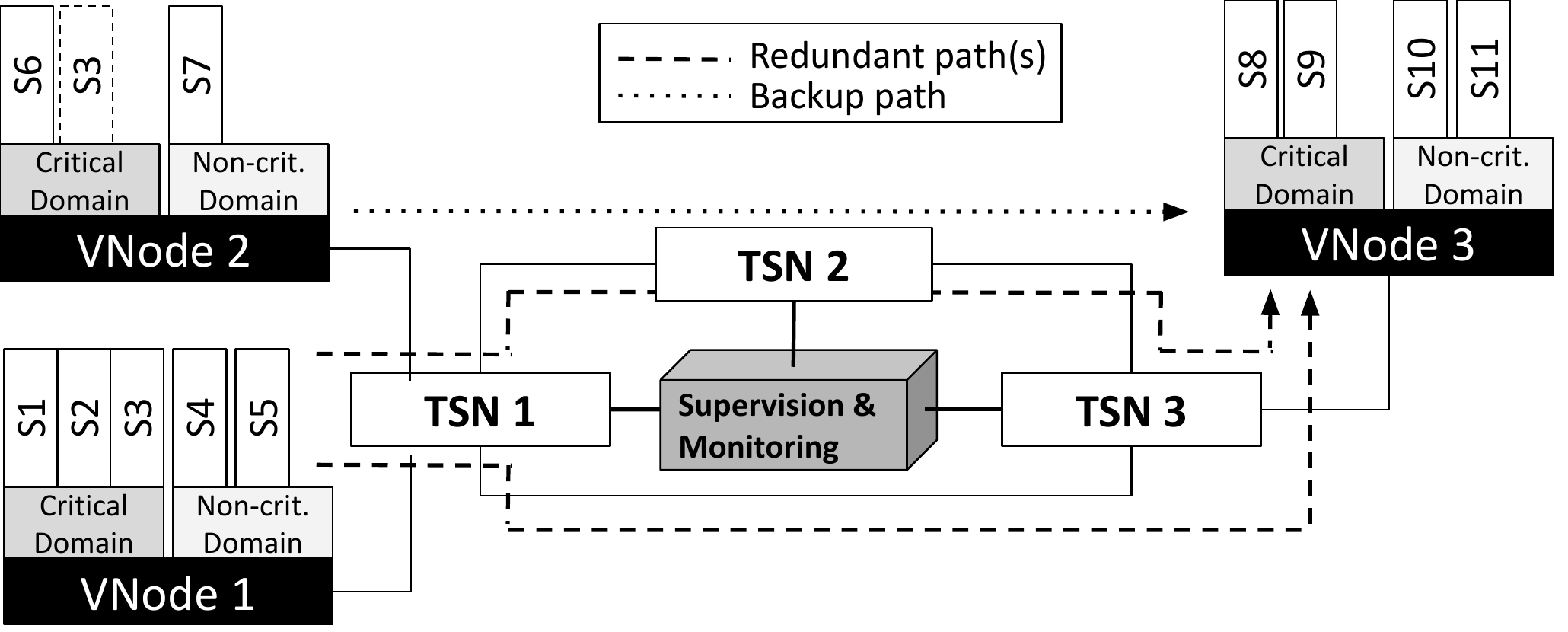}
\caption{A service-oriented MCS with virtualized nodes and TSN bridges.}
\label{fig:architecture}
\end{figure}

In our model, a service-oriented MCS consists of virtualized nodes~(VNodes) inter-connected with TSN bridges that enable TSN protocols for time-sensitive communication. Fig.~\ref{fig:architecture} illustrates an example architecture.
In the figure, each VNode hosts a critical and a non-critical domain to isolate mixed-criticality services from each other. They are connected via TSN bridges~(TSN1-TSN3). 
The services~(S1-S11) are distributed to VNodes by a supervisor~(in the middle) regarding their resource requirements. This supervisor can also configure TSN bridges to establish interconnectivity between services. 
For instance, for connecting critical services S3 and S9 (between VNode 1 and 3), it could reserve sufficient bandwidth and schedule time-sensitive streams over TSN1-TSN2-TSN3 and TSN1-TSN3, where the latter is a redundant path against link failures. 
At the same, non-critical services might communicate without strict quality of services~(QoS) requirements. 
For that, TSN protocols enable the coexistence of mixed-criticality streams on unified networking equipment and establish proper resource provision for end-to-end communication.

Moreover, the service-based design allows dynamic reconfiguration of services in case of node failures. For example, in Fig.~\ref{fig:architecture}, a redundant S3~(dashed) instance is deployed at VNode 2. In case of a failure at VNode 1, which hosts the primary S3 instance, the redundant one can be activated by the supervisor. Besides, the remaining services at VNode 1 could also be migrated to other nodes with available resources. A monitoring component~(in the middle, integrated into the supervisor) is required to detect such incidents and assist the supervisor in taking necessary actions.

\subsection{System Components} \label{sec:components}
We design and implement two main components in the system architecture shown in Fig.~\ref{sec:architecture}: Supervisor and VNode. 

\begin{enumerate}[leftmargin=*]
\item \textbf{Virtual Node~(VNode)} hosts two isolated domains over a hypervisor for critical and non-critical services. These domains are maintained by the \textit{domain manager}. Each domain further hosts their virtual services in containers and has a \textit{container manager} for maintaining services.  
\begin{itemize} 
\item \textbf{Domain Manager~(DM)} implements functions to create, remove, and monitor domains and tracks the resource usage for the respective VNode. It further communicates with the container manager(s) of both domains to regulate service-related operations.
\item \textbf{Container Manager~(CM)} provides an interface to create, remove, deploy, and execute services as virtual containers. It is also responsible for the status monitoring of those services.
\end{itemize}

Note that VNodes utilize nested virtualization, i.e., virtual domains also host virtual containers using different virtualization technologies described later.

\item \textbf{Supervisor} commands to VNodes to create, remove, and migrate the domains and containers, i.e., services, and thus is responsible for service supervision over the network. It consists of three modules.
\begin{itemize}
\item \textbf{Controller} provides the interface to access to VNodes in the system to perform domain and service-related operations remotely. It regularly collects heartbeats from VNodes, including their local status, e.g., resource usage and status of hosted services.
\item \textbf{Distribution Manager~(DTM)} takes the required system services regarding their resource and intercommunication requirements as input, DTM decides on (i) at which VNodes the respective services can be hosted and (ii) on which network paths they can communicate. The supervisor then uses the controller interface to configure VNodes according to this decision. 
\item \textbf{Failure Manager~(FM)} compares the local status of VNodes gathered by the controller with the global system information of the supervisor. If it detects an inconsistency, e.g., an offline domain or failed service, it specifies the required service migrations and reinitiations to ensure service availability. The supervisor then reconfigures VNodes accordingly.
\end{itemize}
\end{enumerate}

We use Xen hypervisor\footnote{Xen Project, https://xenproject.org/} to host virtual domains over VNodes since it is open-source and already takes place in critical domains. Furthermore, we use open-source Linux containers~(LXC)\footnote{Linux Containers, https://linuxcontainers.org/} for the virtual container that encapsulates the services as it provides a minimum overhead. 
We implemented all managers from scratch using Python v3.5.

\subsection{Interconnectivity}

The supervisor collects periodical heartbeats from VNodes. For this, the controller module maintains remote procedure call~(RPC) channels to the DMs of each VNode. RPC provides dedicated, unicast, and confidential channels that can be easily tracked. Using the same channel, it also sends all domain or service-related requests, such as atomic domain or service creating and removing calls or multiple requests to perform a service migration. We use gRPC\footnote{gRPC, https://grpc.io/} to implement these channels.

Apart from that, VNodes communicate over TSN bridges using TSN protocols. Among several protocols, IEEE 802.1CB Frame Replication and Elimination for Redundancy~(FRER)~\cite{ieee-802.1cb} enables redundant communication via disjoint paths against link failures, similar to shown in Fig.~\ref{fig:architecture}. Since we focus on fault tolerance scenarios, we configure FRER for inter-service communication. Moreover, the monitoring component analyzes time-sensitive traffic to ensure the configured and deterministic communication behavior. 

\section{Prototype and Demonstration} \label{sec:prototype}
This section describes our prototype and demonstration scenarios that utilize the components in Section~\ref{sec:components}.

\subsection{Prototype Setup}
\begin{figure}
	\includegraphics[width=\linewidth]{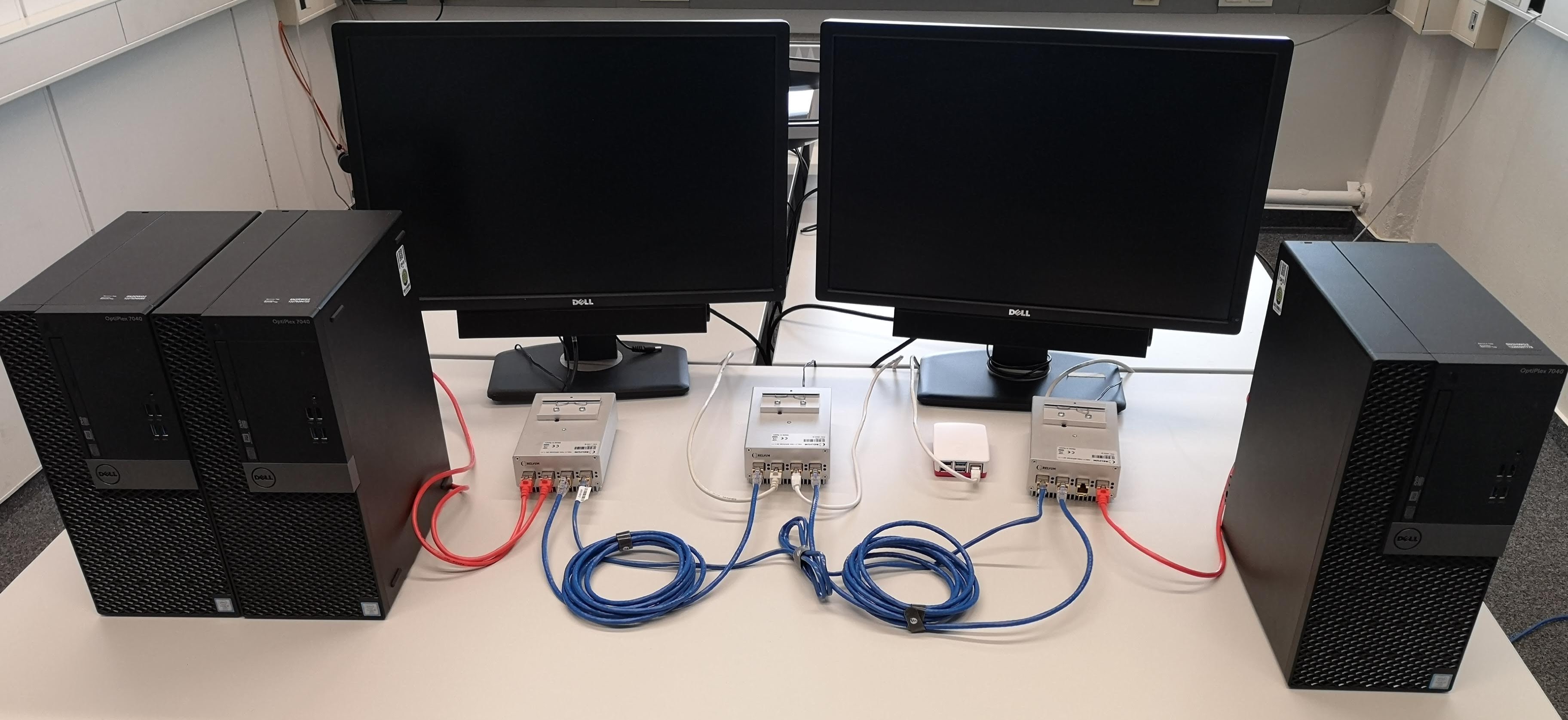}
	\caption{The demonstrator prototype.}
	\label{fig:prototype}
\vspace*{-0.5cm}
\end{figure}

Fig.~\ref{fig:prototype} shows the components in our prototype. It prototype consists of (i) three RELYUM TSN bridges\footnote{RELYUM TSN Bridge, https://www.relyum.com/web/rely-tsn-bridge/} connected in a ring topology, (ii) three virtualized DELL OptiPlex Mini towers as VNodes, and (iii) a DELL T140 server as a supervisor and monitoring component, and (iv) a Raspberry Pi board as an attacker node for the security-related scenarios. The fourth server is not visible in the photo but is roughly similar to the servers shown. The red and blue cables are for VNode-to-bridge and bridge-to-bridge connections, respectively. The supervisor and the attacker node are connected via grey cables.
 We set up the same topology shown in Fig.~\ref{fig:architecture}. Two VNodes (VNode 1 and 2) are connected to the first TSN bridge~(TSN1, on the left in Fig.~\ref{sec:prototype}), and another one (VNode 3) is attached to the third TSN bridge~(TSN3, on the right in Fig.~\ref{fig:prototype}). 

The supervisor is connected to the second bridge~(TSN2, at the middle in Fig.~\ref{fig:prototype}). Besides, a TSN monitoring module~(TSNM) is deployed beside the supervisor on the same server. We developed TSNM in our previous work~\cite{ergenc22tsnzeek} to monitor selected TSN protocols, including FRER, and detect TSN-specific security threats. Note that it is not an internal component of the supervisor but an independent module. All nodes also have an attached screen as a part of the demonstration (two of them excluded from the figure for a better view).

\subsection{Demonstration Scenarios} \label{scenarios}
In our prototype, we demonstrate four scenarios regarding (i) automatic service distribution, (ii) service migration against node failures, (iii) redundancy against link failures, and (iv) intrusion detection for TSN traffic.

\begin{enumerate}[leftmargin=*]
\item \textbf{Optimal service distribution:} The system initially requires the deployment of virtual services on VNodes. In the first scenario, the supervisor automatically distributes video streaming and receiving services to VNodes. It uses our previous optimal service distribution model presented in~\cite{ergenc2020dsn} or a manual configuration file within the DTM. It considers limited VNode resources in terms of their CPU and memory, and the resource requirements of the respective services for the distribution. As a result, VNode 1 streams a flight instruction video, e.g., assuming the system represents a flight cabin network,  and VNode 3 receives this stream. The communication between these VNodes is established using IEEE 802.1 FRER over two redundant paths, i.e., TSN3-TSN2-TSN1 and TSN3-TSN1. All bridges are manually configured with the MAC addresses of VNodes for demonstration purposes. In the end, we observe the video displayed on the screen attached to VNode 3. Meanwhile, the TSNM logs the FRER traffic passing through TSN2 shown on its respective screen.
\item \textbf{Fault tolerance against link failures:} We then demonstrate the benefits of  FRER against link failures. We plug off the Ethernet cable between TSN3 and TSN2 to mimic a broken link on the path TSN3-TSN2-TSN1. It does not affect the video streaming, e.g., an interruption or a delay, since FRER ensures seamless redundancy over the path TSN3-TSN1. However, the TSNM can no longer log the FRER traffic since the respective path is unavailable due to the link failure.
\item \textbf{Fault tolerance against node failures:} In the third scenario, we disconnect VNode3 from the network to demonstrate a node failure. As a result, the supervisor cannot receive periodical heartbeats from VNode 3 and triggers failover via its FM. The FM utilizes a fault-tolerance heuristic, i.e., selecting a backup VNode with the highest available resources, to migrate the video receiving service from VNode 3 to VNode 2 in real time. After a short delay for starting new containers and deploying the respective service on VNode 2, we observe the flight instruction video continues from when VNode 3 fails on the display attached to VNode 2. We can also follow the TSNM logs with an alternated MAC address. 
\item \textbf{Intrusion detection for TSN traffic:} The last scenario demonstrates the detection of TSN-specific attacks via the monitoring module, TSNM. We connect an attacker node to TSN2 and conduct several FRER-related attacks listed in~\cite{ergenc21tsn}. As a result, we observe that the TNSM detects malicious attempts and raises alerts displayed on the screen.
\end{enumerate}

\section{Conclusion} \label{sec:conclusion}

Service-oriented architecture~(SOA) and IEEE 802.1 Time-sensitive Networking~(TSN) are two recent paradigms that ease designing mission-critical systems~(MCSs) that typically consist of dedicated components with static configuration. In this work, we investigate the integration of these emerging technologies to build flexible and resilient MCSs. For this, we implement a prototype with virtualized components and a supervisor connected with TSN bridges. We demonstrate several scenarios to show that our design enhances flexibility and adaptability of MCSs to recover them in case of potential failures via dynamic failovers and redundancy mechanisms. Additionally, we show that our supervisor and monitoring module detect service failures and TSN-specific security threats.

\balance
\bibliographystyle{ieeetr}
\bibliography{IEEEabrv,references}

\end{document}